\documentclass[acus]{jac2000} 
\usepackage{epsfig}
\def\gappeq{\mathrel{ \rlap{\raise.5ex\hbox{$>$}}
                      {\lower.5ex\hbox{$\sim$}}  } }
\def\lappeq{\mathrel{ \rlap{\raise.5ex\hbox{$<$}}
                      {\lower.5ex\hbox{$\sim$}}  } }
\begin{document}
\title{Ground Motion Studies and Modeling for the 
Interaction Region \\ of a Linear Collider
\thanks{Work supported by the U.S. Department of Energy, 
Contact Number DE-AC03-76SF00515.}}
\author{A.~Seryi, M.~Breidenbach, J.~Frisch \\
{\it Stanford Linear Accelerator Center, Stanford University, 
Stanford, California 94309 USA}
}
\maketitle
\begin{abstract}
Ground motion may be a limiting factor in 
the performance of future linear colliders. 
Cultural noise sources, an important component of 
ground motion, are discussed here, with data from
the SLD region at SLAC.
\end{abstract}

\section{Introduction}

Ground motion may be a limiting factor in the performance of 
future linear colliders because it causes continuous misalignment of the 
focusing and accelerating elements.
Understanding the ground motion, including finding driving mechanisms 
for the motion, studying the dependence on  
geology, local engineering, etc., and creating ground motion 
models that permit evaluation of the collider 
performance, are essential for the optimization of the 
linear collider.

There has been a lot of progress in understanding motion of 
the ground and its modeling in recent years, which has allowed us to 
build both general and specific ground motion 
models for a particular location. For example, the 
model presented in \cite{slacmodel} includes 
systematic, diffusive, and fast motion based on 
various measurements performed at the SLAC site. 

However, several important features are not sufficiently 
well studied and consequently are not yet adequately 
represented in this (or other) models or in the underlying 
analytical approach. 
Proper representation of cultural noise
is a major concern. The model mentioned above  
is based on measurements of the fast motion performed 
at night in sector 10 of the SLAC linac \cite{ZDR}, 
one of the quietest locations at SLAC. 
The corresponding model of the correlation 
is suitable for the case when the noise sources 
are located remotely from the points of observation.
Cultural noise may not only increase 
the fast frequency power spectrum, for example 
as shown in Fig.\ref{slac1}, 
but also the correlation model  
may have to be changed if the noise sources are located 
in the vicinity of or between the points of interest. 
Cultural noise sources, located above or 
inside the tunnel, can locally increase the amplitudes 
of motion. The model, and the analytical framework, 
however, assume that the spectrum of motion 
or the correlation do not depend on location, 
which is natural for the spectral approach
based on the use of the 2-D spectrum $P(\omega,k)$,
which cannot depend on position.
This issue should be handled by use of a 
local addition $p(\omega,s)$ to the spectrum 
which would describe (together with corresponding 
correlation information) each noise source located 
in the vicinity. Here $\omega=2\pi f$, $f$ -- frequency, 
$k$ -- wavenumber, $s$ -- position. See \cite{sn} for 
more detailed definitions. 
In some cases, a function $\psi(\omega,s)$ 
which would characterize local amplification of vibrations, 
for example due to the resonant properties of girders, 
should also be used. 

Cultural noise in the detector area of a linear collider 
is of special concern. The most severe position tolerances are 
for the final quadrupoles. Various systems of the 
detector and the detector hall will unavoidably alter 
the natural ``quietness'' of the area.

Studies of vibration noise have recently been performed
in the HERA Hall East \cite{montag}. The observed motion at HERA
was found to be quite large, 
for example the rms motion above 1~Hz 
reachs 100--200~nm. This high level of vibrations 
at HERA appears to be caused by the high urbanization 
of the area. 

In the studies presented below noise in the SLAC Large 
Detector has been investigated.

\section{Noise in SLD detector area}

\begin{figure}[b]
\vspace{.13cm}
\centering
{\vbox{
\epsfig{file=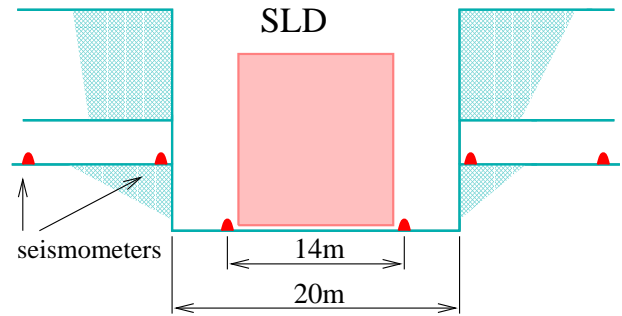,height=0.99\columnwidth,angle=-90}
}}
\vspace{-.1cm}
\caption{Schematics of SLD area showing location 
of seismoprobes installed on the floor of the pit and in 
final focus tunnels.  
The doors of the detector, the superconducting triplets and 
probes installed on them are not shown. }
\label{pit1}
\end{figure}

Vibrations studies are currently being performed in the SLD pit    
at SLAC. The SLD detector is shutdown and 
represents an ideal test bench for such studies. 
Eight seismoprobes have been installed in the detector 
area. Two broadband Streckeisen STS-2 seismometers are placed
under the detector on the concrete floor with 14~m separation  
between them as shown in Fig.\ref{pit1}. 
Four Mark~L4 geophones are placed in the final 
focus tunnels, and two 
piezosensors on the superconducting triplet and 
on the detector itself. The complete results of these studies
in the SLD hall will be reported elsewhere \cite{sm}. 
We present here only the 
results for the floor vibration under the detector.

\begin{figure}[th]
\vspace{.13cm}
\centering
{\vbox{
\epsfig{file=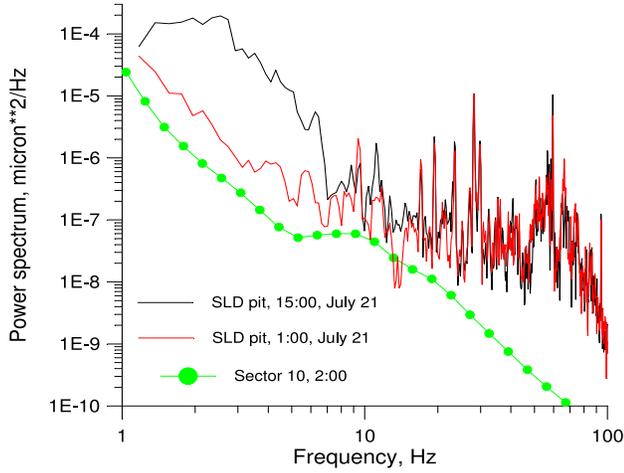,height=0.99\columnwidth,width=0.75\columnwidth,angle=-90}
}}
\vspace{-.1cm}
\caption{Power spectra measured at 2a.m in SLAC 
sector 10 \cite{ZDR} compared with spectra 
measured by the STS-2 probe placed on the concrete floor 
under the South door of the SLD detector.}
\vspace{-.12cm}
\label{slac1}
\end{figure}

\begin{figure}[t]
\vspace{.13cm}
\centering
{\vbox{
\epsfig{file=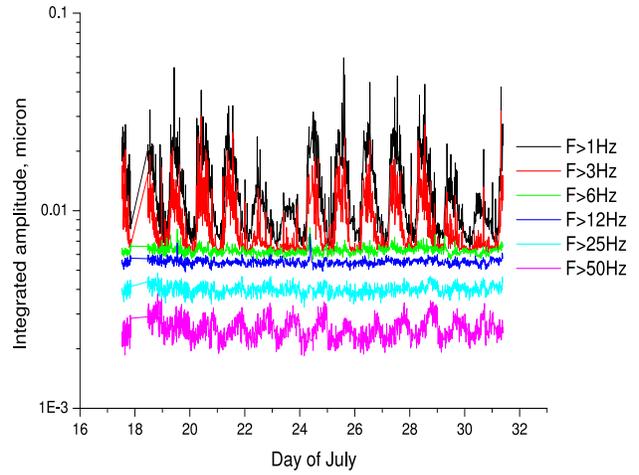,height=0.99\columnwidth,width=0.75\columnwidth,angle=90}
}}
\vspace{-.4cm}
\caption{Rms amplitudes in different frequency bands 
measured by the STS-2 probe placed on the concrete floor 
under the North door of the SLD detector in July 2000. }
\vspace{-.125cm}
\label{sts2}
\end{figure}

\begin{figure}[t]
\vspace{.2cm}
\centering
{\vbox{
\epsfig{file=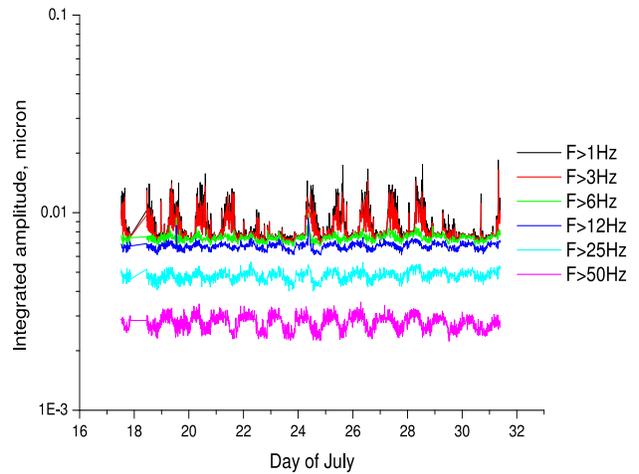,height=0.99\columnwidth,width=0.75\columnwidth,angle=90}
}}
\vspace{-.4cm}
\caption{Rms amplitudes in different frequency bands 
of the difference of displacement measured by two 
STS-2 probes placed with 14~m separation 
on the concrete floor under the South and the 
North doors of the SLD detector.}
\vspace{-.2cm}
\label{diff}
\end{figure}

The power spectra measured by the STS-2 probes during day and night 
is shown in Fig.\ref{slac1}. The high frequency part of 
the spectra ($F \gappeq 10$~Hz) is clearly much noisier than 
that measured 
in sector 10. However, the day-night variation is absent, 
which means that this noise is produced mostly by local 
(in the SLD building and nearby vicinity) sources, while 
the contribution from more remote sources (traffic, etc.) 
is much less pronounced. 
On the other hand, the low frequency contribution of the traffic and 
other cultural noises produced inside and outside of SLAC
has clear day-night variations, as seen in Fig.\ref{slac1},\ref{sts2}
and \ref{diff}.

\begin{figure}[h]
\vspace{0.23cm}
\centering
{\vbox{
\epsfig{file=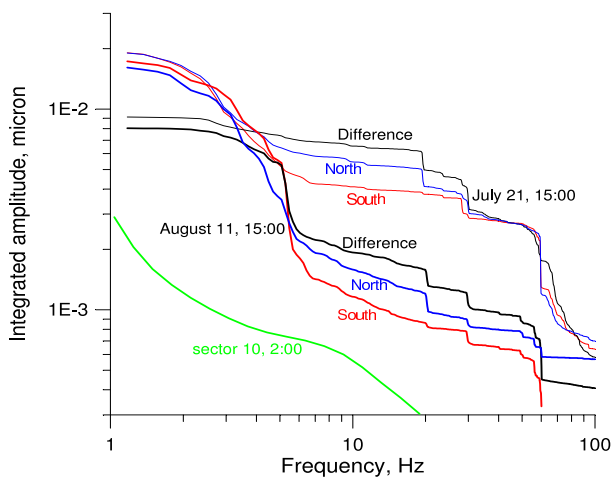,height=0.97\columnwidth,width=0.75\columnwidth,angle=-90}
}}
\vspace{0.23cm}
\caption{Integrated spectrum (amplitude for $F>F_0$) 
corresponded to measurements in SLAC sector 10 at 2:00 
compared with the spectra measured by probes 
placed under the SLD detector with 14~m separation 
at 15:00 on July 21 and at 15:00 on August 11;  
most of the noise sources in the building turned off 
at this later date. }
\vspace{-.05cm}
\label{integ}
\end{figure}

\begin{figure}[t]
\vspace{.13cm}
\centering
{\vbox{
\epsfig{file=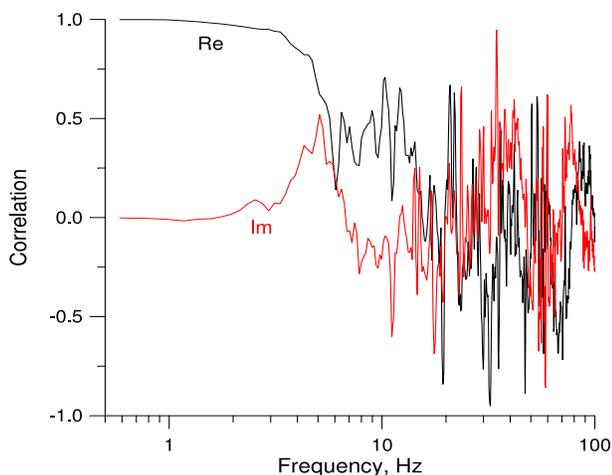,height=0.97\columnwidth,width=0.75\columnwidth,angle=-90}
}}
\vspace{-.3cm}
\caption{Correlation (real and imaginary parts) 
of the motion measured by the STS-2 probes 
placed under the SLD detector with 14~m separation 
on August 11. Averaged over 123 files with 30 seconds 
record length. }
\vspace{-.5cm}
\label{corr}
\end{figure}

The integrated amplitude 
shown in Fig.\ref{sts2},\ref{diff} and \ref{integ} is 
defined as the integral over the power spectrum from a specific
frequency $F$ to a maximal 
frequency: 
\\[-3mm]
$$
\mathrm{rms} = \left( \int\limits_{F}^{F_{max}} p(f) df \right)^{1/2} .
$$
\\[-3mm]
One should note that the rms amplitude of the difference of 
displacements of two points, in the case where these motions are 
uncorrelated, can exceed each of the individual rms values
(for example by a factor of $\sqrt{2}$ if these two rms values 
are equal). The motion under the South and North part of the 
SLD detector, shown in Fig.\ref{integ}, is mostly uncorrelated for 
frequencies higher than about 4~Hz, as seen in Fig.\ref{corr},  
though at some particular frequencies 
the correlation is noticably nonzero
even for $f \gappeq 4$~Hz. 
For identical probes the imaginary part of 
the correlation must be zero if the power spectra
in the two places are equal \cite{sn}. One can see in Fig.\ref{integ} 
that these power spectra are, in fact, quite different 
and so the imaginary part of the correlation shown in 
Fig.\ref{corr} is essentially nonzero.

The measurements presented 
in Fig.\ref{slac1}, \ref{sts2}
and \ref{diff} were performed when 
most of the SLD electronics was on (with its local ventilation) 
and the building ventilation operating. 
The water flow in the SLD conventional solenoid was set to 
one third of the nominal level, approximately 300 gallons per minute. 
The floor motion was found to be greatly influenced by 
the ventilation system of the building 
(located in the North part of the SLD hall) and, to a lesser extent, 
by the SLC and SLD water pumps located about 20~m 
North of the building.

The NLC, operating at a repetition frequency of 120~Hz, 
will be sensitive to the jitter of its final focusing doublet 
at frequencies above approximately 6~Hz (beam-based feedback 
can presumably take care of the beam offsets below this frequency). 
As we see from the Fig.\ref{diff}, even without any significant 
precautions to reduce the noise, the difference of the 
floor motion measured by the two STS-2 probes separated by 14~m 
is about 8~nm for $F \gappeq 6$~Hz, which is 
roughly twice the typical NLC vertical beam size. 

By turning off most (but still not all) of the equipment, including 
the SLD and SLC water pumps, the building ventilation 
and most of electronics (which would require  
proper engineering of these subsystems for NLC) 
this difference can be decreased to about 2~nm \cite{sm}. 
As we see in Fig.\ref{integ}, even in this case, the North 
probe, located closer to the noisier North part of 
the building, shows larger vibrations. 
Therefore, further reduction of the difference value 
would seem to still be possible.

Of course the motion of the final quadrupoles 
cannot be as low as the motion 
of the floor because the supports cannot be made ideally rigid. 
The strategy we consider involves active stabilization of the final 
quadrupoles by using inertial sensors possibly 
in combination with an optical reference to the ground. 
In one of the proposals \cite{bowden} the optical path 
would pass from the final quadrupoles through 
the detector to a common location under the detector. 
This has the disadvantage of putting significant constraints
on the detector design. Such a configuration of the detector 
is now considered unlikely to be necessary. 

However, if the optical reference is desired 
in addition to the inertial sensors to improve the performance 
of the inertial stabilization, the optical reference 
can be made to the floor (possibly to local pits) under each
of the final quadrupoles (approximately at the same 
positions where the STS-2 probes were placed in our measurements). 
The necessary correction of the differential motion of the floor 
could then be done by using seismometers located at these reference 
locations. One can see that in the conditions similar to those of 
the SLD area, 
where the spectrum of motion drops quite rapidly with frequency, 
this strategy would work even without significant additional 
engineering for noise reduction. 

The optical path for the reference to the ground could be located 
outside of the detector, greatly simplifying its design 
and operation. The newly designed final focus system \cite{newffs}, 
which allows a doubling of $L^*$, would simplify the 
detector design even further. 

One can see that the SLD area, after proper engineering, 
or a site with similar characteristics, would be compatible 
with a linear collider having nanometer scale beam sizes.

\section{Conclusion}

Several aspects of ground motion 
require particular attention, namely 
studies and modeling of cultural noises and 
in particular those generated in the detector 
area of a linear collider. 
Studies of the cultural noise 
in the SLAC SLD area presented in this paper 
will help to determine the engineering requirements 
of various subsystems of the detector 
to be compatible
with NLC requirements. 

We would like to thank G.Bowden and T.Raubenheimer 
for various useful discussions.

\end{document}